\documentclass{appolb}
\usepackage{epsfig}

\begin{document}
\eqsec  
\title{Dark matter in dwarf galaxies of the Local Group
}
\author{Ewa L. {\L}okas
\address{Nicolaus Copernicus Astronomical Center, Bartycka 18, 00-716 Warsaw, Poland}
}
\maketitle
\begin{abstract}
We review basic properties of the population of dwarf galaxies in the Local Group focusing on dwarf spheroidal
galaxies found in the immediate vicinity of the Milky Way. The evidence for dark matter in these objects is
critically assessed. We describe the methods of dynamical modelling of such objects, using a few examples of the
best-studied dwarfs and discuss the sources of uncertainties in mass estimates. We conclude with perspectives
for dwarf galaxies as targets for dark matter detection experiments.
\end{abstract}

\section{Introduction}

The population of dwarf galaxies in the Local Group offers a unique opportunity to test our theories of structure
formation in the Universe. Starting from the Magellanic Clouds which were known since antiquity, the census of
the dwarf galaxies in our immediate vicinity still grows. The sample of dwarf galaxies in the Local Group can be
divided using the morphological criteria into classes of dwarf irregulars (dIrr) that are flattened, rotating,
bright and still
forming stars and dwarf spheroidals (dSph) which are rounder, faint and contain mostly old stellar populations
dominated by random motions.

Dwarf galaxies of the Local Group tend to cluster around the two main hosts, the Milky Way and Andromeda, exhibiting a
pronounced morphology-density relation: while dSphs are typically found close to one of the big galaxies, dIrrs
occupy more isolated regions. The origin of this relation is an interesting issue that the theories of structure
formation must address. Another question in which dwarf galaxies played a role is the so-called problem of missing
satellites: the number of observed satellites of the Milky Way is much smaller than predicted by theories based on cold
dark matter. Dwarf galaxies, especially nearby dSphs, have also drawn attention once a significant number of stellar
velocities could be measured and their masses could be determined. These masses turned out to be much larger than
could be explained by the stellar content indicating large amounts of dark matter present.

\begin{figure}
\epsfig{file=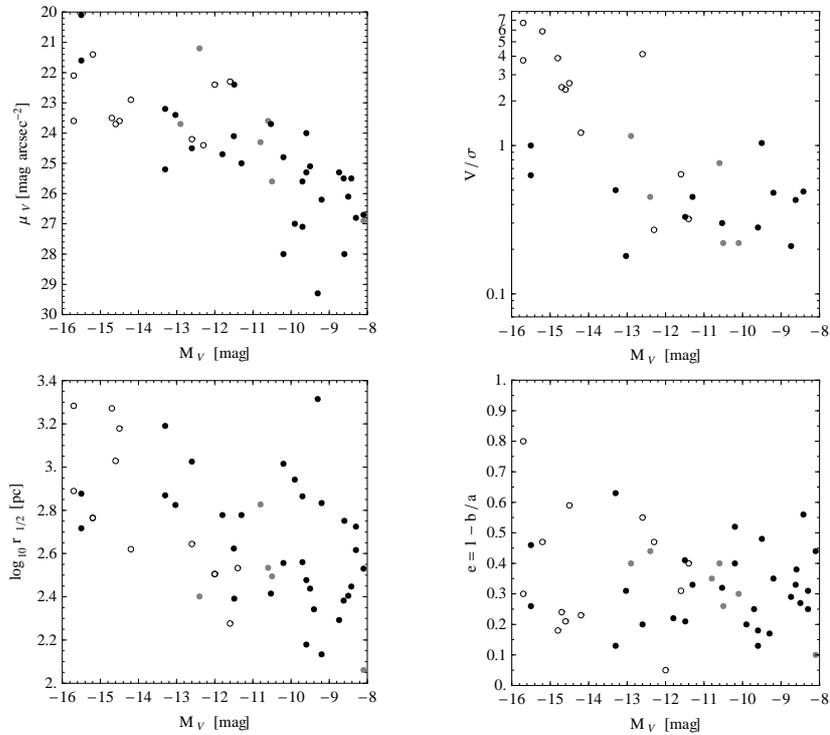, height=10cm}
\caption{Observational parameters of dwarf galaxies in the Local Group:
the central surface brightness $\mu_V$ (upper left panel), the half-light radius $r_{1/2}$ (lower left panel),
the ratio of the rotation velocity to the velocity dispersion $V/\sigma$ (upper right panel)
and the ellipticity $e=1-b/a$ (lower right panel) as a function of the total visual magnitude $M_V$.
The open circles show the data for dIrr galaxies,
filled black circles the data for the dSph and dSph/dE galaxies and gray circles for transitory dIrr/dSph dwarfs.}
\label{lokas_fig1}
\end{figure}

Observational properties of dwarf galaxies \cite{mateo98} are usually characterized by the total
visual magnitude $M_V$, the central
surface brightness $\mu_V$, the half-light radius (containing half the total luminosity) $r_{1/2}$,
the ellipticity $e=1-b/a$ and the ratio of the rotation velocity to the velocity dispersion $V/\sigma$. The
latter four quantities are plotted in Figure~\ref{lokas_fig1} as a function of magnitude for the dwarf galaxies in
the Local Group with $-16$ mag $< M_V < -8$ mag using data compiled in \cite{lokas11}. The sample includes the eight
``classical" best-known dSph galaxies, satellites of the Milky Way: Carina, Draco, Fornax, Leo I, Leo II, Sculptor,
Sextans and Ursa Minor with magnitudes in the range $-13$ mag $< M_V < -8$ mag which all have low central
surface brightness $\mu_V$ ($>22.4$ mag arcsec$^{-2}$) and small half-light radii ($< 0.7$ kpc). The properties
of these eight dwarfs are listed in Table~\ref{properties}. All classical dwarfs
are believed to be characterized by large mass-to-light ratios between 10 and a few hundred solar units.

\begin{table}
\begin{center}
\caption{Properties of the classical dSph galaxies, satellites of the Milky Way.}
\begin{tabular}{lrcccc}
\hline
Dwarf	     &  $M_V \ $ & $r_{1/2}$ & $\mu_V$             & $V/\sigma$ & $e=1-b/a$   \\
galaxy       &  [mag] &  [kpc]    & [mag arcsec$^{-2}$] &            &             \\
\hline
Carina       &$	-8.62 $ & 0.241 &  25.5	& 0.43	&	0.33     \\
Draco        &$	-8.74 $ & 0.196 &  25.3 & 0.21	&	0.29     \\
Fornax       &$	-13.03$ & 0.668 &  23.4 & 0.18	&	0.31     \\
Leo I        &$	-11.49$ & 0.246 &  22.4 & 0.33  &       0.21     \\
Leo II       &$	-9.60 $ & 0.151 &  24.0	& 0.28	&	0.13     \\
Sculptor     &$	-10.53$ & 0.260 &  23.7 & 0.30	&	0.32     \\
Sextans      &$	-9.20 $ & 0.682 &  26.2 & 0.48	&	0.35     \\
Ursa Minor   &$	-8.42 $ & 0.280 &  25.5	& 0.49	&	0.56     \\
\hline
\label{properties}
\end{tabular}
\end{center}
\end{table}

In recent years we have witnessed discoveries of new dSph galaxies in the Local Group, mainly in the northern part
of the Sloan Digital Sky Survey \cite{belokurov07}.
These new dwarfs are generally fainter and more irregular in shape than the classical ones but
not more distant. If analogous discoveries follow in the southern hemisphere, they may significantly help to solve
the problem of missing satellites.
Spectroscopic studies of the stellar populations \cite{simongeha} of the faint dwarfs revealed quite large velocity
dispersions and thus large masses for these dwarfs were estimated suggesting even higher dark matter content than in
well-known dSphs. It has been speculated \cite{strigari} that the characteristic masses of all dwarfs are of the
order of $10^7$ solar masses, independently of their luminosity. A more acceptable proposal \cite{walker}
suggests instead a universal relation between the half-light radius and the mass contained within it.

\section{Evidence for dark matter}

\begin{figure}
\begin{center}
\epsfig{file=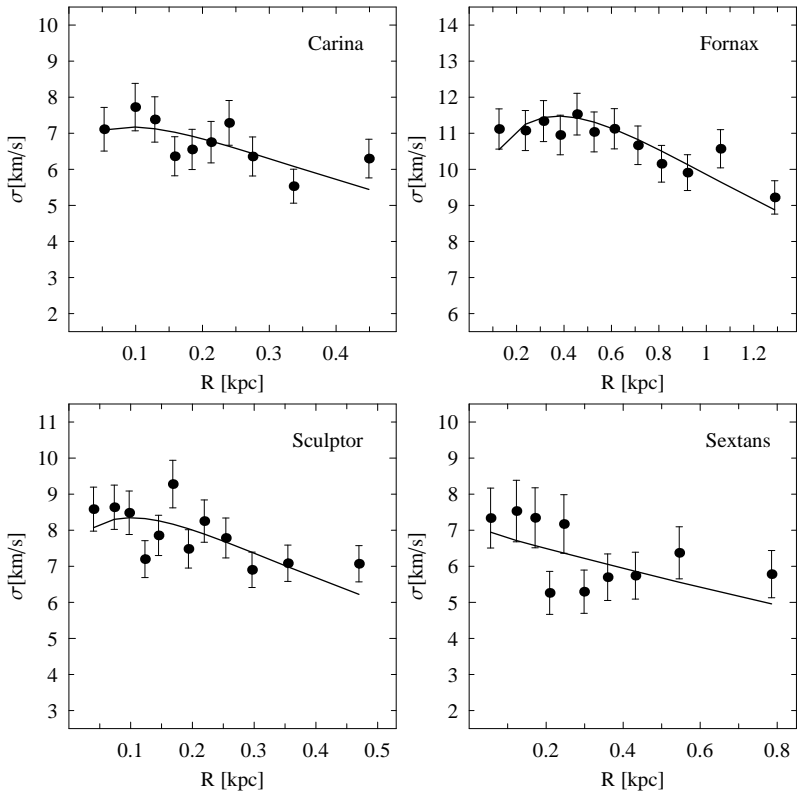, height=11cm}
\caption{Velocity dispersion profiles for the member stars in the four dSph galaxies: Carina, Fornax, Sculptor and
Sextans. Errors indicate the
sampling errors of the dispersion. Black lines show the best-fitting profiles from fitting velocity dispersion
profiles with the solutions of the Jeans equation.}
\end{center}
\label{lokas_fig2}
\end{figure}

Dynamical modelling of dSph galaxies relies on measuring the velocity distribution of stars in a galaxy and modelling
it using methods based on the virial theorem or its extensions. The most common approach is to use the Jeans
equation \cite{lokas02}
which relates the velocity dispersion measured in the galaxy to the underlying gravitational potential.
Currently available data comprise a few hundred up to more than two thousand stellar velocities measured in a single
dSph galaxy which allows us to study the velocity dispersion profile as well as higher velocity moments \cite{lokas09}.
Examples of such profiles are shown in Figure~\ref{lokas_fig2} for the four classical dwarfs with most numerous
data sets available at present \cite{walker09a}.

Fitting the solutions of the Jeans equation to the measured velocity dispersion profiles allows us to constrain the
properties of the mass content of the galaxy. If we assume that mass follows light, the problem can be reduced to
fitting two parameters: the total mass $M$ and the parameter of the velocity anisotropy of stellar orbits $\beta$.
For the four dwarf galaxies from Figure~\ref{lokas_fig2} such modelling yields constraints on these two parameters
shown in Figure~3 with stellar orbits remarkably close
to isotropy ($\beta=0$). The best-fitting masses translate to rather large mass-to-light ratios of the order of
10 solar units for Fornax and Sculptor and
above 50 for Carina and Sextans \cite{lokas09}, signifying the presence of large amounts of dark matter.

\begin{figure}
\begin{center}
\epsfig{file=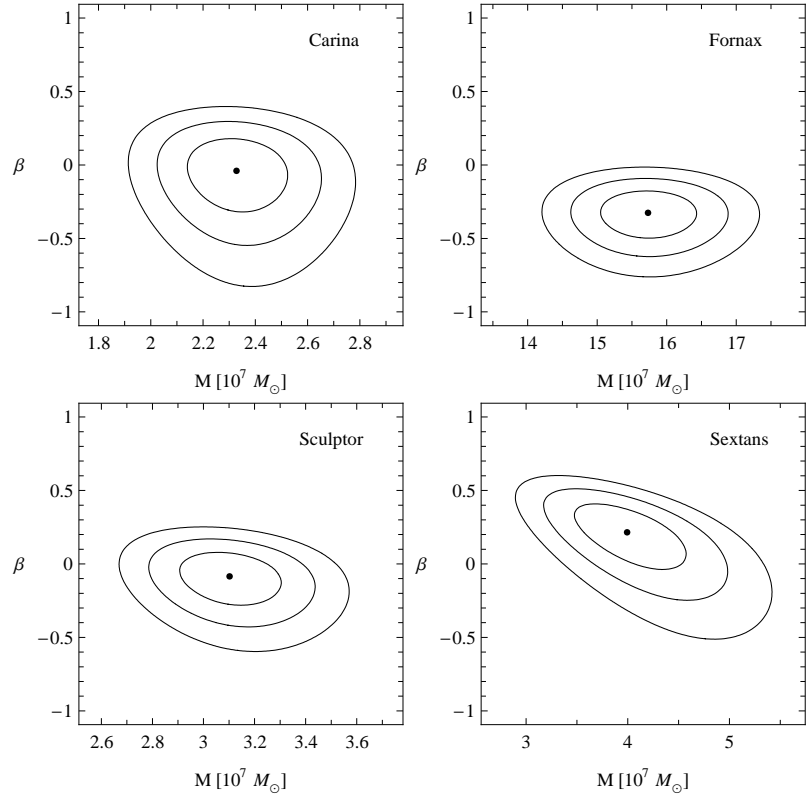, height=11cm}
\caption{The $1\sigma$, $2\sigma$ and $3\sigma$ probability contours showing constraints on the two fitted
parameters: the total mass $M$ and velocity anisotropy $\beta$. The constraints follow from the sampling errors of
the measured velocity dispersion profile. Black dots indicate the best-fitting parameters.}
\end{center}
\label{constraints}
\end{figure}

The reliability of such inferences has been a subject of a long, on-going debate. One popular objection often raised is
that Jeans modelling rests on the assumption that the galaxies are in equilibrium, while if they are strongly
affected by tidal forces from the Milky Way, this may not be the case. Second, even if the objects are
essentially self-gravitating
and in equilibrium, the kinematic samples used for the modelling may be contaminated by tidally stripped stars
counted as members. Third, binary stars present in the population may boost the measured velocity dispersion.
In addition, the most popular assumptions in the modelling: that the objects are spherical and $\beta$ is independent
of radius (or zero everywhere) may be too simplistic.

Of all these objections, the contamination by tidally stripped stars seems to be the most serious issue. Using
$N$-body simulations it has been demonstrated \cite{klim07, klim09, kazan11} that dwarf galaxies orbiting the
Milky Way retain their identity and remain in relative equilibrium for billions of years in spite of being
heavily stripped of stars and dark matter. The stripped material forms extended tidal arms, which, if pointing
towards the observer, may significantly contaminate the samples increasing the measured velocity dispersion.
Efficient procedures to deal with this contamination have been developed \cite{klim07} and it has been demonstrated
on artificial data generated from the simulations \cite{lokas08} that once the data are cleaned of such
interlopers, the mass and anisotropy estimates based on Jeans formalism and its extensions are in good agreement
with true values.

In particular, the velocity dispersion profiles shown in Figure~\ref{lokas_fig2} were obtained
from data preselected using such procedures. Note, that both the measured and the fitted velocity dispersion
profiles in the Figure are decreasing with the projected radius from the centre of the dwarf $R$, as expected for
self-gravitating galaxies, and in contrast to contaminated samples where the dispersion tends to increase at
large distances.

The best-fitting solutions of the Jeans equation shown in Figure~\ref{lokas_fig2} were found with the assumption
that mass follows light. This assumption is justified for dwarfs heavily affected by tides which tend to strip the
(initially more extended) dark halo more effectively than the stars. However, certainly not all satellites of
the Milky Way fall into this category, a notable example being the Draco dSph, where the models with mass following
light fail to reproduce the data in a satisfactory way. If the parameters of the dark matter halo are allowed to
vary, the modelling based on the Jeans equation alone is plagued by a strong degeneracy between the shape of the halo
and the anisotropy parameter. In particular, the same dispersion profile can be reproduced by galaxy models with
tangential orbits ($\beta<0$) and compact dark haloes or radial orbits ($\beta>0$) and extended haloes.

A way to overcome this difficulty is to extend the Jeans formalism to include higher (fourth order) velocity
moment (the kurtosis) which is sensitive mainly to anisotropy. Once the anisotropy is constrained by kurtosis, the
density profile of the dark halo can be constrained as well, so fitting both moments simultaneously allows us to
estimate both properties. This procedure has been successfully applied to the Draco dwarf {\cite{lmp05} giving
a very high mass-to-light estimate of around 300 solar units for this dwarf.

Unfortunately, including higher velocity moments does not help to constrain the inner slope of the dark matter
distribution in dSph galaxies. For example, it has been demonstrated for the Draco dwarf that the cuspy and a core-like
dark matter halo fits the available data equally well \cite{sanchezconde07}.
Measurements of this property are important because they may help to solve the so-called
cusp/core problem in dwarf galaxies, a tension between theoretical models which predict steep inner slopes and
observations of rotationally supported dwarfs which tend to be reproduced better by cores in dark matter distribution.

\section{Conclusions}

The case for a high dark matter content in dSph galaxies of the Local Group is quite strong. The objects are
found to contain significant amounts of dark matter even if the modelling is performed with very conservative
assumptions such as the one that mass follows light and on samples which were subject to restrictive algorithms
of interloper rejection. It has been demonstrated that the classical dwarfs, even if strongly affected by
tidal force from the Milky Way can be reliably modelled by standard methods assuming virial equilibrium.

The high mass-to-light ratios of the classical dwarfs may turn out to be even higher in the case of the newly
discovered population of ultra-faint satellites of the Milky Way. If the trend is confirmed, the fainter dwarfs
may become excellent targets for experiments aiming for the direct detection of dark matter in the Universe via
self-annihilation of dark matter particles.
Dwarf galaxies are good candidates for such targets not only because of the high dark matter content but also
because they have low astrophysical gamma ray backgrounds and they are small and localized so may easily fit
in the field of view of many instruments.

Recently, observations of dwarf galaxies with the Fermi Large Area Telescope began to yield interesting
constraints on the parameters of a variety of supersymmetric models that provide candidates for weakly
interacting massive particles as dark matter \cite{abdo10}. Although no clear signal of annihilation
has been detected from dwarf galaxies so far, upper limits on the annihilation cross-section can be obtained.
The upper limits do not yet reach the interesting region of density parameter values compatible with the
standard cosmological model, but are already able to exclude a significant number of supersymmetric models of
particle physics.

\section*{Acknowledgment}

This research and the author's participation in the XXXVth International Conference of Theoretical Physics
``Matter to the Deepest: Recent Developments in Physics of Fundamental Interactions"
was partially supported by the Polish National Science Centre under grant N N203 580940.


\begin{thebibliography}{99.}

\bibitem{mateo98} M. L. Mateo, {\em ARA\&A\/}, {\bf 36}, 435 (1998).

\bibitem{lokas11} E. L. {\L}okas, S. Kazantzidis, L. Mayer, {\em ApJ\/}, {\bf 739}, 46 (2011).

\bibitem{belokurov07} V. Belokurov et al., {\em ApJ\/}, {\bf 654}, 897 (2007).

\bibitem{simongeha} J. Simon, M. Geha, {\em ApJ\/}, {\bf 670}, 313 (2007).

\bibitem{strigari} L. E. Strigari, J. S. Bullock, M. Kaplinghat, J. D. Simon, M. Geha, B. Willman, M. G. Walker,
         {\em Nature\/}, {\bf 454}, 1096 (2008).

\bibitem{walker} M. G. Walker, M. Mateo, E. W. Olszewski, J. Pe\~narrubia, N. W. Evans, G. Gilmore,
	{\em ApJ\/}, {\bf 704}, 1274 (2009).

\bibitem{lokas02} E. L. {\L}okas, {\em MNRAS\/}, {\bf 333}, 697 (2002).

\bibitem{lokas09} E. L. {\L}okas, {\em MNRAS\/}, {\bf 394}, L102 (2009).

\bibitem{walker09a} M. G. Walker, M. Mateo, E. W. Olszewski, {\em AJ\/}, {\bf 137}, 3100 (2009).

\bibitem{klim07} J. Klimentowski, E. L. {\L}okas, S. Kazantzidis, F. Prada, L. Mayer, G. A. Mamon,
	{\em MNRAS\/}, {\bf 378}, 353 (2007).

\bibitem{klim09} J. Klimentowski, E. L. {\L}okas, S. Kazantzidis, L. Mayer, G. A. Mamon,
	{\em MNRAS\/}, {\bf 397}, 2015 (2009).

\bibitem{kazan11} S. Kazantzidis, E. L. {\L}okas, S. Callegari, L. Mayer, L. A. Moustakas, {\em ApJ\/},
	{\bf 726}, 98 (2011).

\bibitem{lokas08} E. L. {\L}okas, J. Klimentowski, S. Kazantzidis, L. Mayer,
       {\em MNRAS\/}, {\bf 390}, 625 (2008).

\bibitem{lmp05} E. L. {\L}okas, E. L., G. A. Mamon, F. Prada, {\em MNRAS\/}, {\bf 363}, 918, (2005).

\bibitem{sanchezconde07} M. A. Sanchez-Conde, F. Prada, E. L. {\L}okas, M. E. Gomez, R. Wojtak, M. Moles,
	{\em Phys. Rev. D}, {\bf 76}, l23509 (2007).

\bibitem{abdo10} A. A. Abdo et al., {\em ApJ}, {\bf 712}, 147 (2010).

\end{thebibliography}
\end{document}